\documentclass[11pt]{amsart}
\usepackage[centering,a4paper,width=6in,height=9in]{geometry}                
\usepackage[parfill]{parskip}    
\usepackage{graphicx}
\usepackage{amssymb}
\usepackage{epstopdf}
\usepackage{textcomp}
\usepackage{url}
\title{The Recomputation Manifesto}
\author{Ian P. Gent,  12 April 2013\\Version 1: $Revision: 9479 $   }
\address{Ian P. Gent School of Computer Science, University of St Andrews, Fife, KY16 9SX, Scotland, UK.  ian.gent@st-andrews.ac.uk. \url{http://www.cs.st-andrews.ac.uk/~ipg}}

\begin{document}
\maketitle
\renewcommand{\labelenumi}{\arabic{enumi}.}

\newcommand{\manifestoA}{Computational experiments should be recomputable for all time}
\newcommand\manifestoB{Recomputation of recomputable experiments should be very easy}
\newcommand\manifestoC{It should be easier to make  experiments recomputable than not to}
\newcommand\manifestoD{Tools and repositories can help recomputation  become standard}
\newcommand\manifestoE{The \emph{only} way to ensure recomputability is to provide virtual machines}
\newcommand\manifestoF{Runtime performance is a secondary issue}

\begin{quote}
\emph{
\begin{enumerate}
\item \manifestoA
\item \manifestoB
\item \manifestoD 
\item \manifestoC
\item \manifestoE
\item \manifestoF
\end{enumerate}
}
\end{quote}

Replication of scientific experiments is  critical to the advance of science.\footnote{As well as being critical, it is a timely issue, witness this in \emph{Nature}, 3 April 2013 {\cite{worthdoingtwice}.}}
Unfortunately, the discipline of  Computer Science has never treated replication seriously, even though computers are  very good at doing the same thing over and over again.    Not only are experiments rarely replicated, they are rarely
even replicable in a meaningful way.  Scientists are being encouraged to make their source code available \cite{InceD2012Case}, but this is only a small step.
Even in the happy event that  source code can be built and run successfully, 
running code is a long way away from being able to replicate the experiment that code was used for.
 
 I propose that the discipline of Computer Science must embrace replication of experiments as standard practice.   I  propose  that the only credible technique to make experiments truly replicable is to provide copies of virtual machines in which the experiments are validated to run.  I propose that  tools and  repositories should be made available to make this happen. I propose to be one of those who makes it happen. 

I am using the word `recomputation' to mean the replication of computational experiments.
This is the `recomputation manifesto', not the `replication manifesto'. The word  `replication' 
has a well-established technical meaning in computing. 
I am adding a technical meaning to an existing word,\footnote{The word `recomputation' is older than the USA, reported by the  \emph{Oxford English Dictionary} in 1766. It does have some technical meanings, e.g. \cite{DBLP:books/sp/Schulte02}, but none  as widespread as those for `replication'.}
but the new meaning adds only a nuance and also serves usefully to distinguish replication of computational experiments from other types of experiment.

\section*{1. \manifestoA.}

Why recomputable?  And why for all time?

The necessity for recomputation is simple. An experiment  is used to form scientific conclusions, and those conclusions are never final.  They may always be subject to question, so experimental work may need to be repeated.  
I would even argue that the less experiments are recomputed, the \emph{more} important it is that they be recomputable.  The reason is that flaws in an experiment that is frequently recomputed are likely to come to light, but not so if the experiment is rarely recomputed.  The more rarely an experiment is repeated, the longer the time lapse until it is, and the less likely a new researcher will be able to recompute it easily without the original experiment being deliberately made recomputable.  If the original experiment is flawed in some way, misleading results can lie uncorrected in the literature for years \cite{DBLP:conf/csclp/BeckPW04}.

Should experiments be recomputable for all time?  Yes.  Computer Science is unique in having this ability.  Imagine if physicists had the opportunity to look through Galileo's telescope at Jupiter's moons, but had thrown them away on the basis that a description of how to make it was available.   There is no reasonable limit to what is the useful life of an experiment is, so the simplest answer is to have no limit. Moore's law, while it lasts, makes it technically feasible to store experiments forever. 
It's affordable to store experiments: if cost per GB decreases exponentially, a fixed sum is enough to store each GB in perpetuity.  
It is even affordable to rerun all computational experiments in history regularly.    If researchers donated 10\% of 
CPU resources for new experiments to recomputing old ones, all 5 year old experiments could be rerun each year.\footnote{Assuming computing power  doubles every 18 months, CPU resources increase by 10 times every 5 years.}

\section*{2. \manifestoB}

Recomputation should be very easy.  Ideally, very easy means clicking a button.  This will not always be possible, but we should make it as easy  as we possibly can for experimenters to reproduce their own experiments and those of other researchers.   
A lovely example of making it easy to rerun existing experiments is provided by the  website 
\url{runmycode.org},  which experimenters can use to provide a simple interface 
 to test their code on different inputs
\cite{DBLP:conf/eScience/StoddenHP12}.

An important point about this manifesto is that by recomputation I focus mainly on the exact replication of an experiment.  
No other choice is available  since (as I argue below) we have to duplicate the exact experimental conditions to ensure that recomputation is possible.  Drummond argues powerfully that this - which he calls `replicability' - is the poor cousin of  `reproducibility', where experiments are reproduced with changes to factors believed to be insignificant
\cite{drummond09}.   There is no question that key advances in science should be reproducible in Drummond's richer sense, but neither do I accept that  ``the impoverished version, replicability, is one not worth having" \cite{drummond09}. 
To take a non-computational example, it would be wonderful if the original experiments on cold fusion could be replicated exactly.  If there was some technical flaw with the experiments, it could be discovered, or if not the effects that led to the apparent conclusion that cold fusion existed could be investigated.   
Once we have ensured that experiments can be recomputed exactly, we can have the luxury of seeking to ensure that experiments can be reproduced in richer ways. 

\section*{3. \manifestoD}

I draw an analogy with version  control tools.  Learning and using subversion, git, or mercurial adds complication to one's life.  But the benefits are enormous, and without question it is easier overall to build and maintain code using source code control systems than without.    Similarly, repositories such as github or bitbucket greatly enhance development and sharing of code.
Once we have the appropriate tools and repositories we in Computer Science will have an enormous advantage over other sciences.   Many sciences have massive online repositories, of course leveraging advances in Computer Science to their benefit.  We can be unique in having massive online repositories of fully realised experiments, not just the data resulting from experiments.  
We should research and make available tools and repositories for the recomputation of scientific experiments. 
Sadly, tools for recomputation are relatively lacking to date, although there are promising signs.  In a single PhD thesis, Philip Guo introduced Burrito, a tool for electronic lab notebooks, and CDE, an automatic packaging tool for 
experiments in linux \cite{guo-phd}.  Other important recent tools include ``verifiable computational results" \cite{DBLP:journals/cse/GavishD12},  ``Sumatra"  \cite{sumatra-davison} and HAL \cite{DBLP:conf/lion/NellFHL11}.  As well as runmycode.org, other repositories helping to make experiments replicable are \url{myExperiment.org} \cite{DBLP:journals/fgcs/RoureGS09}, 
SHARE \cite{DBLP:journals/sosym/GorpG12}, and the specialist journal ``Image Processing Online" (\url{www.ipol.im}) in which each article is accompanied by runnable code.    

\section*{4. \manifestoC}

How can it be easier to make new experiments recomputable than not to bother?  I draw your attention to Jon  Claerbout's words \cite{claerbout}:
\begin{quote}
\emph{``It's not really for the benefit of other people. Experience shows the principal beneficiary of reproducible research is you the author yourself."} 
\end{quote}
I've been doing major computational experiments for 20 years: some of them I am rather proud of.  But whenever I start on a new set of experiments, my heart sinks a little.  I expect the pain of getting everything set up, and then I expect the difficulty or impossibility of reproducing my experiments if, say, the paper is rejected and I need to rerun them for a later submission.  We should fix this.  Tools should be available to make it easier to run experiments, encourage good experimental practice, and simultaneously make them recomputable.     An experienced experimenter's heart should sink no more than an experienced programmer when starting a new program.    
Just as programmers have comfortable development environments they can use, experimenters should have a range of environments to develop experiments in. 
Repositories of past experiments - one's own and other people's - should act like software libraries to aid rapid development of new experiments.

\section*{5. \manifestoE}

While the other points of the manifesto state how I think the world should be, 
this point is an empirical statement about how the world is.   There are two reasons that virtual machines are - at least for now - the only way to go.  The first is the almost universal (I suspect) experience among anyone who has built software and especially tried to rebuild it later.  A build of the resulting program (and hence experiment) can fail due to arbitrary changes in the machine being used.  This applies even where the machine is physically the same one that was used months ago to run an experiment the first time.    An innocent software update, to a package not obviously used by the build software, can cause disaster.   If the machine is not the  same one or a clone thereof, all bets are off.   The \emph{only} way to ensure an experiment is recomputable is to make available a virtual machine identical to one which was tested and worked when the experiment was originally run. 
The second reason is that the available computers and operating systems change over time.     It may be true that code you make available today can be built with only minor pain by many people on current computers.  That is unlikely to be true in 5 years, and hardly credible in 20.   

A classic example illustrates both these reasons.  SHRDLU is one of the most famous programs in the history of AI \cite{shrdlu}. 
The complete source code is available, but cannot be run.   The physical machine and OS it ran on can be emulated, but we do not have the exact machine state that enabled the program to run.  Amongst other issues, Terry Winograd changed his own copy of Lisp, changes that are now lost \cite{shrdlu-resurrection}.
What we need -- and should provide for new experiments -- is an exact virtual machine in which the experiment worked.
There are isolated cases  of researchers providing virtual machines for recomputation, such as Brown \cite{brown,Brown2012} and SHARE \cite{DBLP:journals/sosym/GorpG12}.  
This should be the standard way Computer Scientists do business.   
Howe makes this case much more extensively, giving 13 reasons why virtual machines in the cloud can improve reproducibility 
\cite{10.1109/MCSE.2012.62}.

This does not mean that all uploads and downloads to repositories should be of full virtual machines.  For ease of use, as well as reducing bandwidth, it's important that tools provide as many ways as possible of allowing experiments to be uploaded and downloaded.    

\section*{ 6. \manifestoF} 

If a measure is deterministic, e.g. the data structure resulting from a deterministic algorithm, it should be identical every time the experiment is recomputed.  
We cannot guarantee the same results in  non-deterministic measures such as cpu time.   
By using virtual machines, we may obtain different results as are seen on a physical machine, and future results on virtual machines may differ from current ones.   Even where serious efforts are made, there is no guarantee that reproducible run-time results  can be obtained.  For example, even though max-clique researchers have a standard methodology for cross-referencing cpu times, Prosser has shown that these results cannot be relied on even approximately
\cite{DBLP:journals/algorithms/Prosser12}.

We must tackle the easier problems first.   Ensuring recomputation of experiments is not easy, but it is vital.  Obtaining meaningfully replicable cpu time comparisons requires significant additional research -- if it is even possible.   By allowing recomputation of experiments, we allow researchers to do them in a variety of environments, discovering if conclusions about run times are universal or contingent. The crucial thing is to preserve scientific experiments. It's unarguable that if we can't recompute an experiment at all, we can't preserve run time performance.

\section*{The recomputation.org Mission}

To help make recomputation a reality,
I am starting \url{http://recomputation.org} as one repository for recomputable experiments, and starting to work on the tools to use on the site.   Based on my arguments above I have the following plans.  The overriding goal is the following mission statement: 
\begin{quote}

\emph{If we can compute your experiment now, anyone can recompute it 20 years from now
}
\end{quote}
The site  \url{recomputation.org} is brand new and (as I write) holds zero experiments.   But the following are the principles which we will use in developing it to deliver the mission statement.
\begin{enumerate}
\item  \url{recomputation.org} will make available virtual machines or equivalent technology to allow exact recomputation of lodged experiments.
\item \url{recomputation.org} will make its best efforts to ensure that all experiments which it believes to be recomputable will remain recomputable for all time. 
\item  \url{recomputation.org} will always be free to those lodging bona-fide scientific experiments and to those obtaining past experiments, provided that: all aspects of the experiments are freely available; 
the experimenter's contributions are open source;  and fees are not charged for the related scientific publication.\footnote{The point of the caveats is that  free service might not be offered to authors who will not make source code available, or to publications such as journals who charge either authors or readers for access.   Equally, 
there is major value in making non-open experiments recomputable too, so if funding is available it may be appropriate to allow these to be freely deposited also.}
\item \url{recomputation.org} will provide its code and tools using an appropriate open source licence, including 
server-side code.

\item  \url{recomputation.org} will serve as a testbed for scientific research into issues such as experimental techniques and methodologies.
\end{enumerate}

\section*{Vote Recomputation}

The recomputation manifesto is similar in spirit to the `Science Code Manifesto' 
 but addresses an orthogonal issue.
The Science Code Manifesto, {\url{sciencecodemanifesto.org}}, demands that code used in scientific publications should be made available in an open way
for the useful lifetime of the publication.   But this is neither a necessary condition for recomputation nor sufficient for it. 
Closed source experiments can be recomputed if the appropriate environment is provided -- e.g. a suitable virtual machine containing the necessary binaries but not source.    Arguably recomputability is \emph{more} important for closed source than for open source.   But it must not be thought that open source is sufficient for recomputability.  The \emph{only} 
guarantee of recomputability is the exact environment being available.
I do endorse the Science Code Manifesto, and initial efforts at  \url{http://recomputation.org} will be focussed on open source efforts only.   As well 
as the scientific desirability of open source, it avoids one form of potential licencing problems in recomputation. 

A manifesto is a call that people reading it should vote for your point of view.   Don't vote with a signature or a petition. Vote by making your computational experiments recomputable.   Do it at \url{http://recomputation.org}, or at your own web site, or at another repository.  But make your experiments recomputable.

\bibliographystyle{plain}

\bibliography{recomp}

\begin{thebibliography}{10}

\bibitem{DBLP:conf/csclp/BeckPW04}
J.~Christopher Beck, Patrick Prosser, and Richard~J. Wallace.
\newblock Trying again to fail-first.
\newblock In Boi Faltings, Adrian Petcu, Fran\c{c}ois Fages, and Francesca
  Rossi, editors, {\em CSCLP}, volume 3419 of {\em Lecture Notes in Computer
  Science}, pages 41--55. Springer, 2004.

\bibitem{brown}
C.~Titus Brown.
\newblock Our approach to replication in computational science.
\newblock \url{http://ivory.idyll.org/blog/replication-i.html}, 2012.

\bibitem{Brown2012}
C~Titus Brown, Adina Howe, Qingpeng Zhang, Alexis~B Pyrkosz, and Timothy~H
  Brom.
\newblock {A Reference-Free Algorithm for Computational Normalization of
  Shotgun Sequencing Data}.
\newblock 2012.

\bibitem{claerbout}
Jon Claerbout.
\newblock Reproducible computational research: A history of hurdles, mostly
  overcome.
\newblock \url{http://sepwww.stanford.edu/sep/jon/reproducible.html}.

\bibitem{shrdlu-resurrection}
Semaphore Corporation.
\newblock {SHRDLU} resurrection.
\newblock \url{http://www.semaphorecorp.com/misc/shrdlu.html}, 2011.

\bibitem{sumatra-davison}
Andrew Davison.
\newblock Automated tracking of computational experiments using sumatra.
\newblock In {\em EuroSciPy: 3rd European meeting on Python in Science}, 2010.

\bibitem{drummond09}
Chris Drummond.
\newblock Replicability is not reproducibility: Nor is it good science.
\newblock {\em Proceedings of the Twenty-Sixth International Conference on
  Machine Learning: Workshop on Evaluation Methods for Machine Learning IV},
  2009.

\bibitem{DBLP:journals/cse/GavishD12}
Matan Gavish and David~L. Donoho.
\newblock Three dream applications of verifiable computational results.
\newblock {\em Computing in Science and Engineering}, 14(4):26--31, 2012.

\bibitem{hownot}
I.P. Gent, S.A. Grant, E.~MacIntyre, P.~Prosser, P.~Shaw, B.M. Smith, and
  T.~Walsh.
\newblock How not to do it.
\newblock {\em Research Report Series-University of Leeds School of Computer
  studies LU SCS RR}, 1997.

\bibitem{DBLP:journals/sosym/GorpG12}
Pieter~Van Gorp and Paul W. P.~J. Grefen.
\newblock Supporting the internet-based evaluation of research software with
  cloud infrastructure.
\newblock {\em Software and System Modeling}, 11(1):11--28, 2012.

\bibitem{guo-phd}
Philip~J. Guo.
\newblock Software tools to facilitate research programming.
\newblock Ph.D. dissertation, Department of Computer Science, Stanford
  University, 2012.

\bibitem{10.1109/MCSE.2012.62}
Bill Howe.
\newblock Virtual appliances, cloud computing, and reproducible research.
\newblock {\em Computing in Science and Engineering}, 14(4):36--41, 2012.

\bibitem{InceD2012Case}
Darrel~C. Ince, Leslie Hatton, and John Graham-Cumming.
\newblock The case for open computer programs.
\newblock {\em Nature}, 482(7386):485--488, february 2012.

\bibitem{DBLP:conf/lion/NellFHL11}
Christopher Nell, Chris Fawcett, Holger~H. Hoos, and Kevin Leyton-Brown.
\newblock Hal: A framework for the automated analysis and design of
  high-performance algorithms.
\newblock In Carlos A.~Coello Coello, editor, {\em LION}, volume 6683 of {\em
  Lecture Notes in Computer Science}, pages 600--615. Springer, 2011.

\bibitem{DBLP:journals/algorithms/Prosser12}
Patrick Prosser.
\newblock Exact algorithms for maximum clique: A computational study.
\newblock {\em Algorithms}, 5(4):545--587, 2012.

\bibitem{DBLP:journals/fgcs/RoureGS09}
David~De Roure, Carole~A. Goble, and Robert Stevens.
\newblock The design and realisation of the my$_{\mbox{experiment}}$ virtual
  research environment for social sharing of workflows.
\newblock {\em Future Generation Comp. Syst.}, 25(5):561--567, 2009.

\bibitem{worthdoingtwice}
Jonathan~F. Russell.
\newblock If a job is worth doing, it is worth doing twice.
\newblock {\em Nature}, 496:7, April 2013.

\bibitem{DBLP:books/sp/Schulte02}
Christian Schulte.
\newblock {\em Programming Constraint Services: High-Level Programming of
  Standard and New Constraint Services}, volume 2302 of {\em Lecture Notes in
  Computer Science}.
\newblock Springer, 2002.

\bibitem{DBLP:conf/eScience/StoddenHP12}
Victoria Stodden, Christophe Hurlin, and Christophe Perignon.
\newblock Runmycode.org: A novel dissemination and collaboration platform for
  executing published computational results.
\newblock In {\em eScience}, pages 1--8. IEEE Computer Society, 2012.

\bibitem{shrdlu}
Terry Winograd.
\newblock Understanding natural language.
\newblock {\em Cognitive Psychology}, 3(1):1--191, 1972.

\end{thebibliography}

\section*{About the Author}

Ian Gent is Professor of Computer Science at the University of St Andrews, Scotland.   
His interest in the proper foundations of empirical science in computing date almost 20 years.  He has given
tutorials  on ``Empirical Methods in CS and AI" at conferences such as IJCAI 2001.
Of his non peer-reviewed papers, his most cited by far is ``How Not To Do It'' \cite{hownot}, a collection of embarrassing mistakes he and colleagues  have made in computational experiments.   To show how good we are at not doing things right,  we mis-spelt the name of one of the authors!  

\section*{Acknowledgements}

I wish to thank many scientists I have discussed these issues with over the years, for example 
my fellow authors of \cite{hownot}. I especially thank  Patrick Prosser for his tenacious pursuit of replication of past experiments and stories of his struggles in achieving them,  Ewan (not Ewen) MacIntyre for accepting our mis-spelling of his name, Adam Barker, and Lars Kotthoff. I also thank more  recent colleagues, including  Edwin Brady, Chris Jefferson, Steve Linton, Ian Miguel, Pete Nightingale,  Karen Petrie, Aaron Quigley, and Jonathan Ward. 

\end{document}